\documentclass[journal,twoside]{IEEEtran} 

\normalsize

\usepackage{cite}
\usepackage{csquotes}

\usepackage{subfigure}
\usepackage{svg}
\usepackage{mwe}
\usepackage{tikz}
\usepackage{tikz-3dplot}
\usepackage{pgfplots}
\pgfplotsset{compat=1.6}
\usetikzlibrary{arrows, decorations.markings}
\usetikzlibrary{calc}                   
\usetikzlibrary{shapes}      
\usetikzlibrary{arrows,decorations.pathmorphing,backgrounds,positioning,fit,petri}

\definecolor{review}{rgb}{0,1,0}
\definecolor{change}{rgb}{1,0,0}
\definecolor{revised}{rgb}{0,0,1}

\definecolor{max}{named}{orange}

\usepackage[nooldvoltagedirection]{circuitikz}
\usepackage{overpic}
\usepackage{todonotes}

\usepackage{amsmath}
\usepackage{siunitx}
\DeclareSIUnit{\fps}{fps}

\usepackage{booktabs}

\begin{document}

\title{On Mutual Information Analysis of Infectious Disease Transmission via Particle Propagation}

\author{Peter~Adam~Hoeher, Martin Damrath, Sunasheer Bhattacharjee, and Max Schurwanz%
\thanks{P.\,A.~Hoeher, M.~Damrath, S.~Bhattacharjee, and M.~Schurwanz are with the Faculty of Engineering, 
Kiel University, Kiel, Germany, e-mail: \{ph,md,sub,masc\}@tf.uni-kiel.de.}
}

\markboth{}%
{}

\IEEEpubid{}

\maketitle

\begin{abstract}
Besides mimicking bio-chemical and multi-scale communication mechanisms, molecular communication forms a theoretical framework for virus infection processes.
Towards this goal, aerosol and droplet transmission has recently been modeled as a multiuser scenario.
In this letter, the ``infection performance'' is evaluated by means of a mutual information analysis, and by an even simpler probabilistic performance measure which is closely related to absorbed viruses.
The so-called infection rate depends on the distribution of the channel input events as well as on the transition probabilities between channel input and output events.  
The infection rate is investigated analytically for five basic discrete memoryless channel models.
Numerical results for the transition probabilities are obtained by Monte Carlo simulations for pathogen-laden particle transmission in four typical indoor environments: two-person office, corridor, classroom, and bus.  
Particle transfer contributed significantly to infectious diseases like SARS-CoV-2 and influenza. 
\end{abstract}


\IEEEpeerreviewmaketitle

\section{Introduction}
\IEEEPARstart{M}{otivated} by Claude E.~Shannon's fundamental model of a noisy transmission system \cite[Fig.~1]{Shannon1948}, viral aerosol information retrieval in communication through exhaled breath has initially been studied in \cite{Khalid2019}.
Respiratory events are exploited as a source message in a human-to-machine communication setup.
After transmission via an atmospheric channel, the message is proposed to be scanned by a nanosensor-based machine-type detector.
Subsequently, in \cite{Khalid2020} the same authors developed a mathematical analysis model of aerosol transmission and detection in order to examine detection probabilities and ranges.
Related work on end-to-end system modeling has been published in \cite{Gulec2020}, but for a human-to-human communication setup, where transmitter/receiver are interpreted as infected/uninfected humans. 
Also, drag and buoyancy are taken into account in \cite{Gulec2020}. 
Inspired by papers such as \cite{Khalid2019,Gulec2020}, the duality between molecular communication and pathogen-laden particle transmission has been explored in \cite{Schurwanz2020a, Schurwanz2020b}.
Specifically, the analogy to a multiuser communication scenario is elaborated.
In \cite{Schurwanz2020a}, it is briefly suggested to use mutual information to measure the ``infection performance'' of a particle-based transmission system.
Possible mutual information minimization techniques to reduce the risk of infection are pointed out in \cite{Schurwanz2020b}.
In the context of infectious disease transmission via particles, novel contributions of the letter include 
\begin{itemize}
    \item an analytical investigation of the mutual information for basic discrete channel models, 
    \item a critical review of a logarithmic infection measure, and 
    \item numerical evaluations of the transition probabilities considering typical environments.
\end{itemize}

In the literature, the distinction between aerosols (small lightweight particles suspended in air) and droplets (larger particles) is not defined uniquely.
Therefore, subsequently we refer to particles as a general expression if size is not relevant. 
Pathogen-laden particles are dubbed infected particles. 

\IEEEpubidadjcol

\section{Notation and Fundamentals}
Subsequently, the end-to-end system is simplified by a discrete memoryless channel (DMC) model.  The validity of this assumption is checked in Section~\ref{sec:numerical}.  Let us denote the events at the channel input by $x_i\in {\cal X}$, $1\leq i\leq L_X$, where ${\cal X}$ is the input alphabet of cardinality $L_X=|{\cal X}|$.  Correspondingly, the channel output events are denoted as $y_j\in {\cal Y}$, $1\leq j\leq L_Y$, where ${\cal Y}$ is the output alphabet of cardinality $L_Y=|{\cal Y}|$.  Channel inputs and outputs are randomly distributed.  The corresponding random variables are $X$ and $Y$, respectively.  $X$ and $Y$ are assumed to be discrete-valued.  Furthermore, let $p_X(x_i)$, $p_Y(y_j)$, $p_{X,Y}(x_i,y_j)$, and $p_{Y|X}(y_j|x_i)$ be the marginal probability mass function of $X$, the marginal probability mass function of $Y$, the joint probability mass function between $X$ and $Y$, and the conditional probability mass function of $Y$ given $X$, respectively.  With these notations, the average mutual information can be expressed as \cite{Cover2006}
\begin{equation}\label{mi}
    I(X;Y) = \sum\limits_i \sum\limits_j p_{X,Y}(x_i, y_j)\cdot \log_2 \frac{p_{X,Y}(x_i, y_j)}{p_X(x_i)\cdot p_Y(y_j)}.
\end{equation}
In terms of the input distribution $p_X(x_i)$ and the transition probabilities $p_{Y|X}(y_j|x_i)$, (\ref{mi}) can be reformulated as 
\begin{eqnarray}
    I(X;Y) &=& \sum\limits_i \sum\limits_j p_X(x_i)\,p_{Y|X}(y_j|x_i)\nonumber\\ 
    && \cdot \log_2 \frac{ p_{Y|X}(y_j|x_i) }{ \sum\limits_\iota p_X(x_\iota)\, p_{Y|X}(y_j|x_\iota) }.
\end{eqnarray}
The transition probabilities are a characteristic of the channel, including receiver-side properties (like the sensitive area), whereas the input distribution is independent of the channel. 
This model separation proves to be quite useful. 
Regarding data transmission, the maximum transmission rate for which a quasi-error-free data transmission is possible, i.e., the channel capacity, is obtained by maximizing the mutual information \cite{Shannon1948, Cover2006}.  
Examples of recent publications considering particle transmission include \cite{Akyildiz2019, Farsad2020}.


Concerning infection, neither the entire mutual information $I(X;Y) = \sum_j I(X;Y=y_j)$ nor $\sum_{j\in {\cal Y}_I} I(X;Y=y_j)$ is relevant, where ${\cal Y}_I$ is the set of infectious output events, but only the individual contributions 
\begin{eqnarray}
    I(X;Y=y_j) &=& \sum\limits_i p_X(x_i)\,p_{Y|X}(y_j|x_i)\nonumber\\ 
    && \cdot \log_2 \frac{ p_{Y|X}(y_j|x_i) }{ \sum\limits_\iota p_X(x_\iota)\, p_{Y|X}(y_j|x_\iota) } 
\end{eqnarray}
for all $j\in {\cal Y}_I$.
$R=I(X;Y\in {\cal Y}_I)$ is dubbed {\em infection rate} and measured in bit/channel event. For $n$ events, the {\em mutual infection} is $n\, R$. 
As opposed to data communication systems, mutual information should be minimized in this context.

\section{Basic DMC Channel Models}
\subsection{Example~1 (Z channel)}
\begin{figure}
\addtolength\abovecaptionskip{-5pt}  
    \centering
    \includegraphics[width=0.9\linewidth]{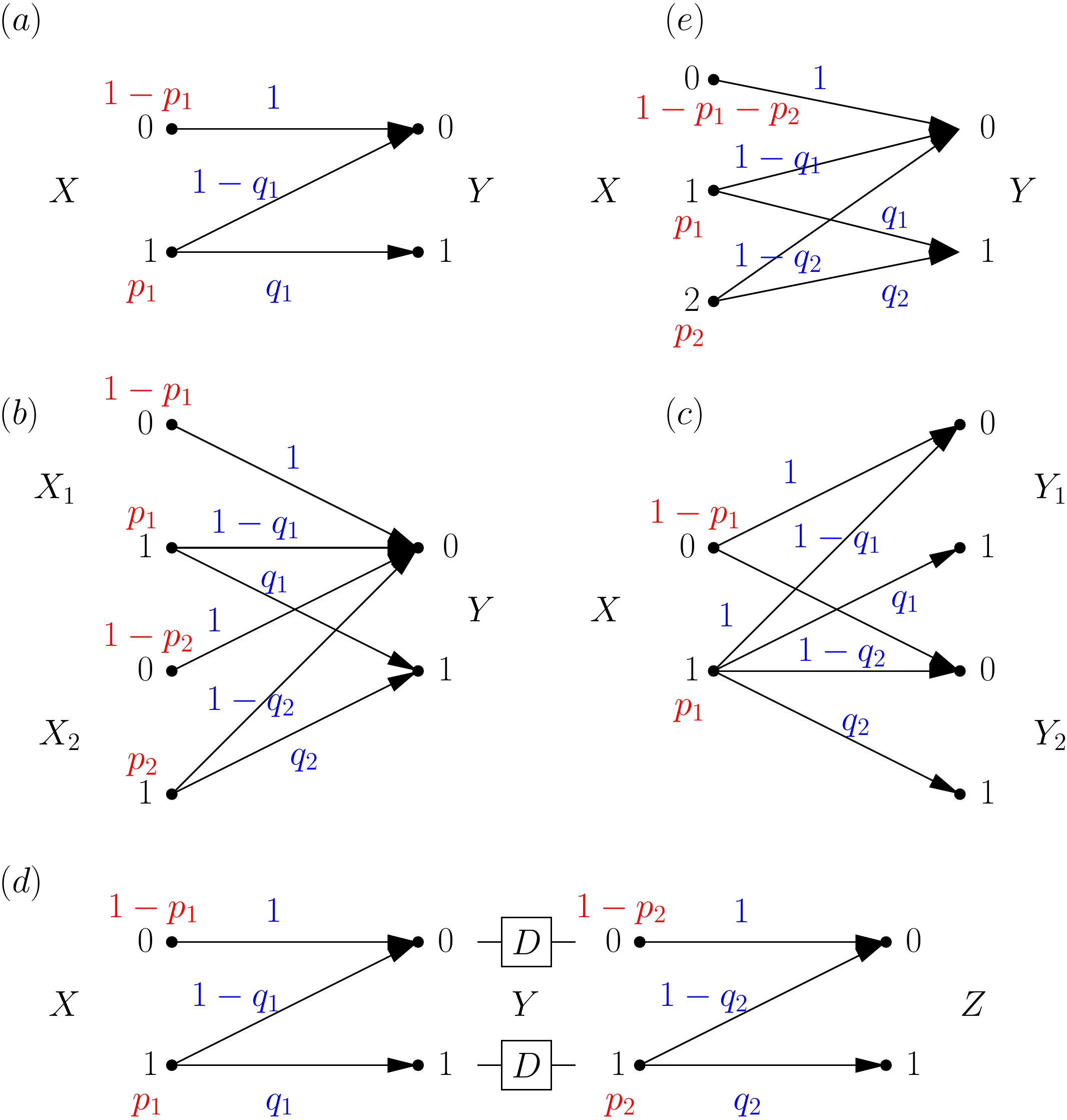}
    \caption{Discrete-time channel models for the five infection scenarios under investigation. 
    Channel input probabilities are marked in red color, whereas transition probabilities are plotted in blue color.}%
    \label{fig:mi_dmc}
    \vspace{-.8em}
\end{figure}

In the first example, the point-to-point scenario shown in Fig.~\ref{fig:mi_dmc}~(a) is considered, where $L_X=L_Y=2$. The two channel input events are labeled ``0'' (referring to non-infected particles) and ``1'' (comprising all infected particles), respectively.  Assume that the emitted particles are infected with probability $p_1$, i.e., $p_X(0)=1-p_1$ and $p_X(1)=p_1$.  The channel output event is labeled ``1'' for absorbed viruses and ``0'' for unabsorbed viruses.  Naturally, the transition probabilities leaving 0 are $p_{Y|X}(0|0)=1$ and $p_{Y|X}(1|0)=0$, because non-infected particles are not able to cause an infection at the receiver side.  Vice versa, the transition probabilities leaving 1 are $p_{Y|X}(0|1)=1-q_1$ and $p_{Y|X}(1|1)=q_1$, respectively.  Only a fraction $q_1$ of infected particles will eventually cause the receiver to become infected.  The only infectious output event is ``1''.  It can be proven that $p_Y(0)=1-p_1\,q_1$ and $p_Y(1)=p_1\,q_1$.
The mutual information between $X$ and $Y=1$ can be written in closed form as
\begin{equation}
    I(X;Y=1) = -q_1\cdot p_1\, \log_2 p_1.
\end{equation}
The maximum is $-(1/e)\log_2(1/e)\cdot q_1$ yielded at $p_1 = 1/e$.  For $0\leq p_1\leq 1/e$, $I(X;Y=1)$ is monotonically increasing, as desired. 
Beyond this maximum, however, the so-called {\em logarithmic infection measure} is meaningless.  We subsequently prove that $p_1\cdot q_1$, called {\em linear infection measure}, is a simpler yet more realistic infection measure.

\subsection{Example 2 (Z channel for two transmitters)}
In the second example, a second transmitter is added, see Fig.~\ref{fig:mi_dmc}~(b).  Now, $L_X=4$ and $L_Y=2$.  The additional input and transition probabilities are called $p_2$ and $q_2$, respectively.
The mutual information between $X=(X_1,X_2)$ and $Y=1$ can be computed as 
\begin{equation}
    I(X;Y=1) = -q_1\cdot p_1\, \log_2 p_1 -q_2\cdot p_2\, \log_2 p_2.
\end{equation}
Note that any additional transmitter in a multipoint-to-point scenario linearly increases the mutual information. 

\subsection{Example 3 (Z channel for two receivers)}
Starting off from Example~1, in the third example a second receiver is added (see Fig.~\ref{fig:mi_dmc}~(c)), i.e., $L_X=2$ and $L_Y=4$.
In this point-to-multipoint scenario the mutual information between $X$ and $Y_1=1$ respectively $Y_2=1$ is yielded as
\begin{eqnarray}
    I(X;Y_1=1) &=& -q_1\cdot p_1\, \log_2 p_1,\nonumber\\
    I(X;Y_2=1) &=& -q_2\cdot p_1\, \log_2 p_1.
\end{eqnarray}
Compared to Scenario~(b), there are less parameters involved because of the common input. 
As opposed to Scenario~(b), the infection rates act individually rather than additive. 

\subsection{Example 4 (Concatenated Z channels, relaying)}
In the fourth example, two Z channels are concatenated as depicted in Fig.~\ref{fig:mi_dmc}~(d).  This scenario emulates relaying with sequential processing. 
Effectively, there are two receivers in this scenario.  The first receiver, $Y$, is infected the same as in Scenario~(a). 
After a certain time delay $D$, receiver $Y$ may turn into a transmitter. 
The second receiver, $Z$, can only be infected by $Y$. 
If $D$ is shorter than the incubation time, $Y$ is not able to amplify the viral load. 
In this case, dubbed passive relaying, the mutual information between $X$ and $Z=1$ is yielded in closed form as 
\begin{equation}
    I(X;Z=1) = -q_2\cdot p_1\, q_1 \log_2(p_1\, q_1).
\end{equation}
According to the data processing theorem, $I(X;Z=1)\leq I(X;Y=1)$.
This is explained by the fact that $p_Y(1)=p_1\cdot q_1\leq p_X(1)$. 
As a result, the relation to the initial particle spreader diminishes with an increasing number of relays. 
For passive relaying, the overall DMC between $X$ and $Z$ is again a Z channel with transition probabilities $p_{Z|X}(0|0)=1$, $p_{Z|X}(0|1)=1-q_1 q_2$, and $p_{Z|X}(1|1)=q_1 q_2$.

If, however, $D$ exceeds the incubation time, the relay acts as an active spreader. 
In that case, 
\begin{equation}
    I(X;Z=1) = -q_2\cdot p'_2 \log_2 p'_2,
\end{equation}
where $p_Y(1)=p'_2$ depends on the efficiency of the relay to boost the viral load. 

\subsection{Example 5 (Nonbinary-input channel)}
A natural extension of the Z channel in Scenario~(a) is shown in Fig.~\ref{fig:mi_dmc}~(e).  In this refinement, the binary input alphabet is replaced by a ternary input alphabet ($L_X=3$). 
Input event~``0'' again corresponds to non-infectious particles, but now input events~``1'' and~``2'' are assigned to aerosols and droplets, respectively.  
The output events are kept unchanged ($L_Y=2$).  Let us assume that emitted aerosols are infected with probability $p_X(1)=p_1$, and emitted droplets with probability $p_X(2)=p_2$.  Accordingly, the transition probabilities causing an infection are denoted as $p_{Y|X}(1|1)=q_1$ and $p_{Y|X}(1|2)=q_2$, respectively.  Since event~0 is not able to cause an infection, $p_{Y|X}(0|0)=1$.  The mutual information between $X$ and $Y=1$ can be computed as 
\begin{equation}
    I(X;Y=1) = -q_1\cdot p_1\,\log_2 p_1 -q_2\cdot p_2\,\log_2 p_2.
\end{equation}
Formally, this result is equivalent to the two-user case in Scenario~(b). 
An extension to multiple particle sizes is apparent. 

\subsection{Medical infection measure}
The number of absorbed viruses can be expressed as 
\begin{equation}\label{V}
    \Phi = n \sum\limits_d p(d)\,q(d)\cdot \eta(d)\,N(d), 
\end{equation}
where $n$ is the number of respiratory events, $p(d)$ and $q(d)$ are the input and transition probabilities as a function of particle size $d$, $N(d)$ the corresponding number of emitted particles per event, and $\eta(d)$ the number of viruses per particle.  A subject is said to be infected if the absorbed viral load $\Phi$ exceeds an infection threshold $\Theta$.  Eqn.~(\ref{V}) in conjunction with the nonbinary-input channel model (e) proves the relevance of the proposed linear infection measure.

\section{Numerical Results for Typical Environments}\label{sec:numerical}
\begin{figure}
\addtolength\abovecaptionskip{-5pt}  
    \centering
    \includegraphics[width=0.9\linewidth]{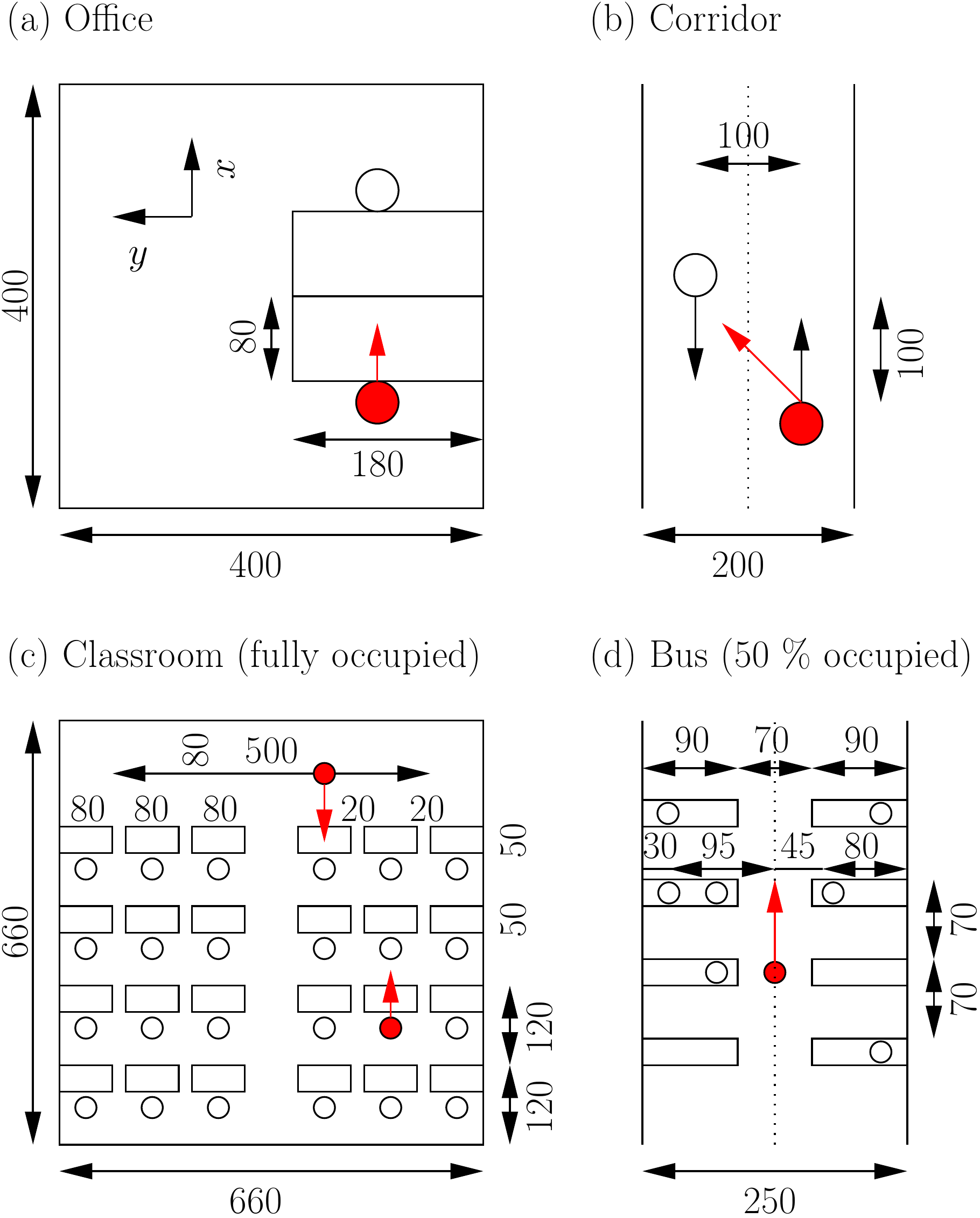}
    \caption{Environments under investigation: (a) office; (b) corridor;  (c) classroom; (d) bus.  Infected users are marked in red.  All dimensions are in cm.}%
    \label{fig:scenarios}
    \vspace{-.8em}
\end{figure}
In this section, transition probabilities will be evaluated for the four environments depicted in Fig.~\ref{fig:scenarios}. 
These environments are interesting from communication theory (point-to-point, point-to-multipoint, and multipoint-to-multipoint scenarios), epidemiology (short-range and long-range transmission), as well as socio-economic aspects. 

Environment~(a) shows a typical office space occupied by two people sitting vis-a-vis in front of their desks for a long period of time, one of whom is believed to be infected. 
A duration of 4~h, for example, corresponds to $n=240$ coughs.  

In Environment~(b), two people meet in a hallway. 
They are looking at each other, but continue walking. 
At a distance of 1~m, the infected person coughs once ($n=1$). 
The non-infected person is walking at a speed of 1~m/s (that is 3.6~km/h) through the emitted particle cloud. 

Environment~(c) features a fully occupied classroom.  
One student as well as the teacher are assumed to be infected. 
The teacher's position is uniformly random distributed in front of the class. 
A lesson typically lasts for 90~min ($n=90$). 

A similar situation is studied in Environment~(d), where a $2\times 2$-seater bus is assumed: although the bus is occupied by just 50~\%, any infected person boarding or leaving the bus is likely to infect surrounding passengers.  
A single cough event is emulated at the position drawn in the figure. 

Random trajectories of exhaled particles are emulated by means of Monte Carlo simulations. 
In all cases, the height $h$ of the mouth is taken to be 1.20~m and 1.64~m for sitting and standing people, respectively. 
One or two infected subjects are considered, coughing once per minute. 
Per cough, 4973 water particles are expected to be released at an initial speed of 11.2~m/s, see \cite{Gulec2020} and reference [37] therein. 
Their diameter $d$ is taken from \cite{Gulec2020} and reference [29] therein, where $d_\mathrm{min}=2~\mu$m and $d_\mathrm{max}=2000~\mu$m. 
Since the size of a COVID-19 virus is approximately 0.1~$\mu$m $\ll d_\mathrm{min}$, the number of viruses encapsulated per water particle is $\sim d^3$. 
For instance, given an oral fluid average viral load of $7\cdot 10^6$ copies per milliliter, the probability that a $20~\mu$m particle contains a virion is about 3~\%.
Since the viral load depends on the course of the disease and varies individually, we have changed the viral load and hence $p_X(1)$ over a wide range. 
The temporal resolution is $\Delta T=10^{-4}$~s, which is sufficiently small given a maximum particle velocity on the order of 10~m/s and a spatial resolution better than of 1~mm. 
The beam width of emitted particles is zero-mean Gaussian distributed ${\cal N}(0, 6.25^{\circ})$ in horizontal and vertical directions \cite{Schurwanz2020a}. 
The mean value of the angle of incidence is also assumed to be Gaussian distributed to account for head rotations: 
${\cal N}(0, 30^{\circ})$ in horizontal direction (except for the hallway, where the mean value is fixed to be $45^{\circ}$ at the time of incidence) and ${\cal N}(0, 10^{\circ})$ in vertical direction (except for the hallway, where the mean value is zero). 
An absorbing receiver with a radius of 5~cm around the depicted position(s) is acquired. 
The number of hits are counted receiver-wise.
In all environments, a ceiling height of 3~m is assumed, except for the bus having an aisle height of 2.3~m. 
Particles hitting the walls, ceiling, or floor are absorbed. 
Gravity and air drag are reproducibly taken into account as
\begin{eqnarray}
v_{x,y}[k+1] &\!\!\!\!=\!\!\!\!& v_{x,y}[k] -\alpha v_{x,y}[k], ~0<\alpha<1, ~k\geq 0,\\
v_z[k+1] &\!\!\!\!=\!\!\!\!& v_z[k] - g \Delta T - \alpha v_z[k], ~g = 9.81~\mathrm{m/s}^2,~~
\end{eqnarray}
where $v_{x,y,z}[0]$ are the initial velocities in the direction of the $x,y,z$ axes.
The air drag is determined by $\alpha=(\beta/m) \Delta T$, where $\beta=3\pi \eta d$ is the Stokes drag coefficient and $\eta\approx 1.85\cdot 10^{-5}$~kg/(m\,s) the dynamical viscosity \cite{Aydin2020}.
Insertion of the mass $m=\rho_\mathrm{water} (\pi/6)\,d^3$ yields $\alpha=18 \eta \Delta T/(\rho_\mathrm{water}\, d^2)$, where $\rho_\mathrm{water}\approx 998-994$~kg/m$^3$ at $20-35^\circ$C. 
Note that $\rho_\mathrm{water}\gg \rho_\mathrm{air}$. 
Neglecting buoyancy, turbulence, droplet shrinking in dry air, and assuming that the particles do not interact with each other, the maximum velocity is $|v_\mathrm{z,\,\infty}|=m\, g/\beta$ if the particles are emitted horizontally. 
Consequently, the channel model is memoryless if the time difference between respiratory events exceeds the fall time $h/|v_\mathrm{z,\,\infty}|$ for most of the pathogen-laden particles. 
%

\begin{figure}
\addtolength\abovecaptionskip{-5pt}  
    \centering
    \subfigure[Office room]{\includegraphics[width=0.49\linewidth]{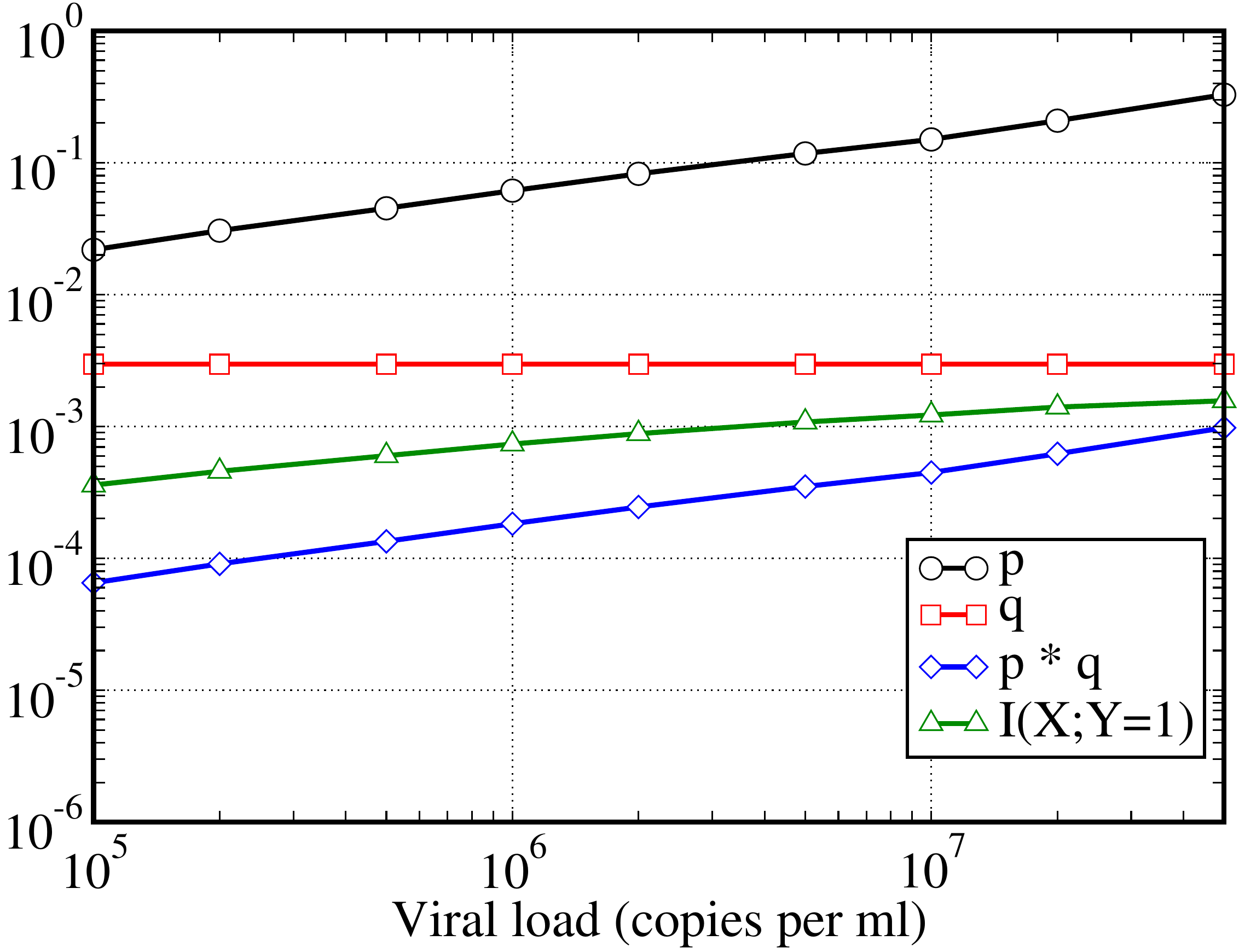}}
    \subfigure[Corridor]{\includegraphics[width=0.49\linewidth]{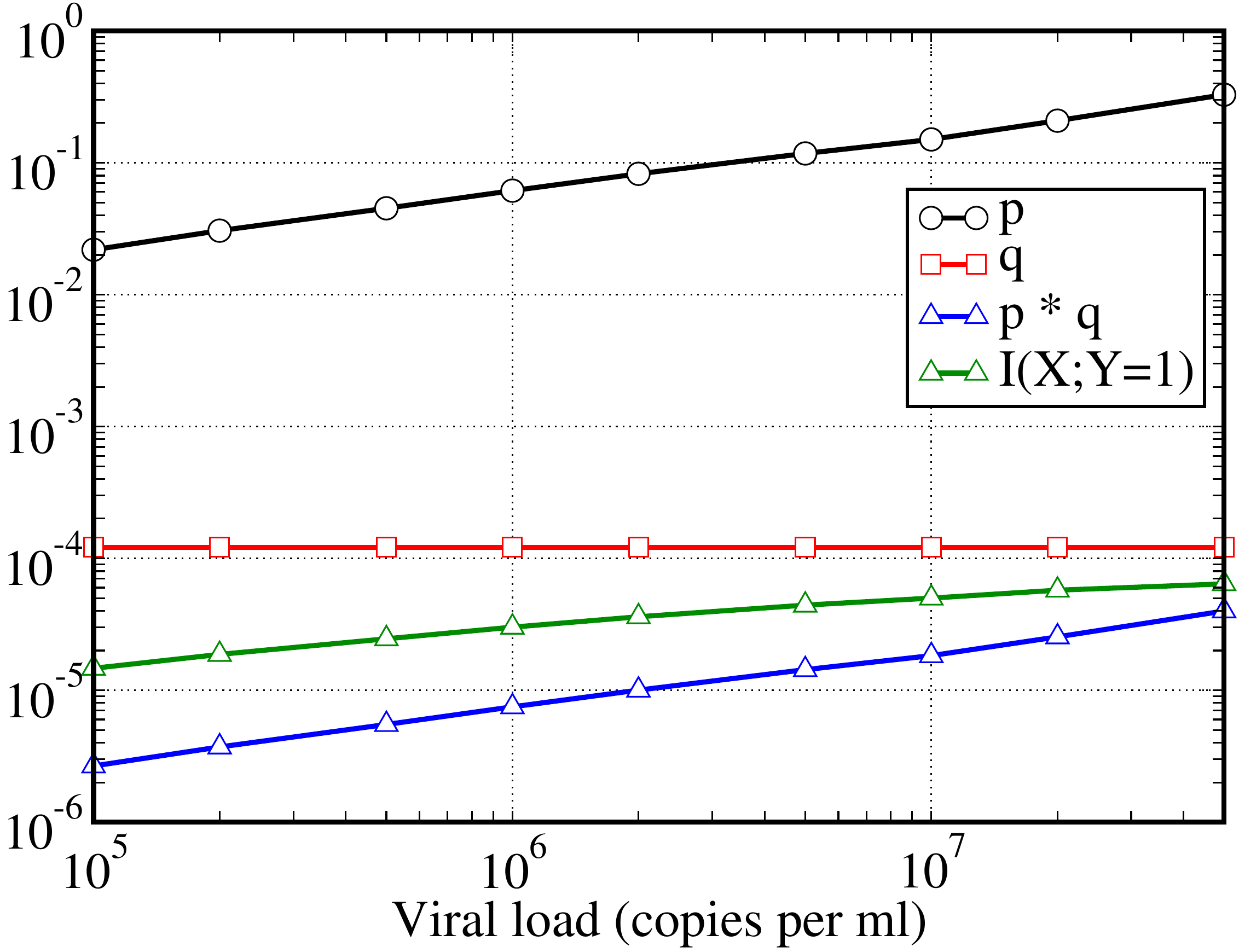}}
    \subfigure[Classroom]{\includegraphics[width=0.49\linewidth]{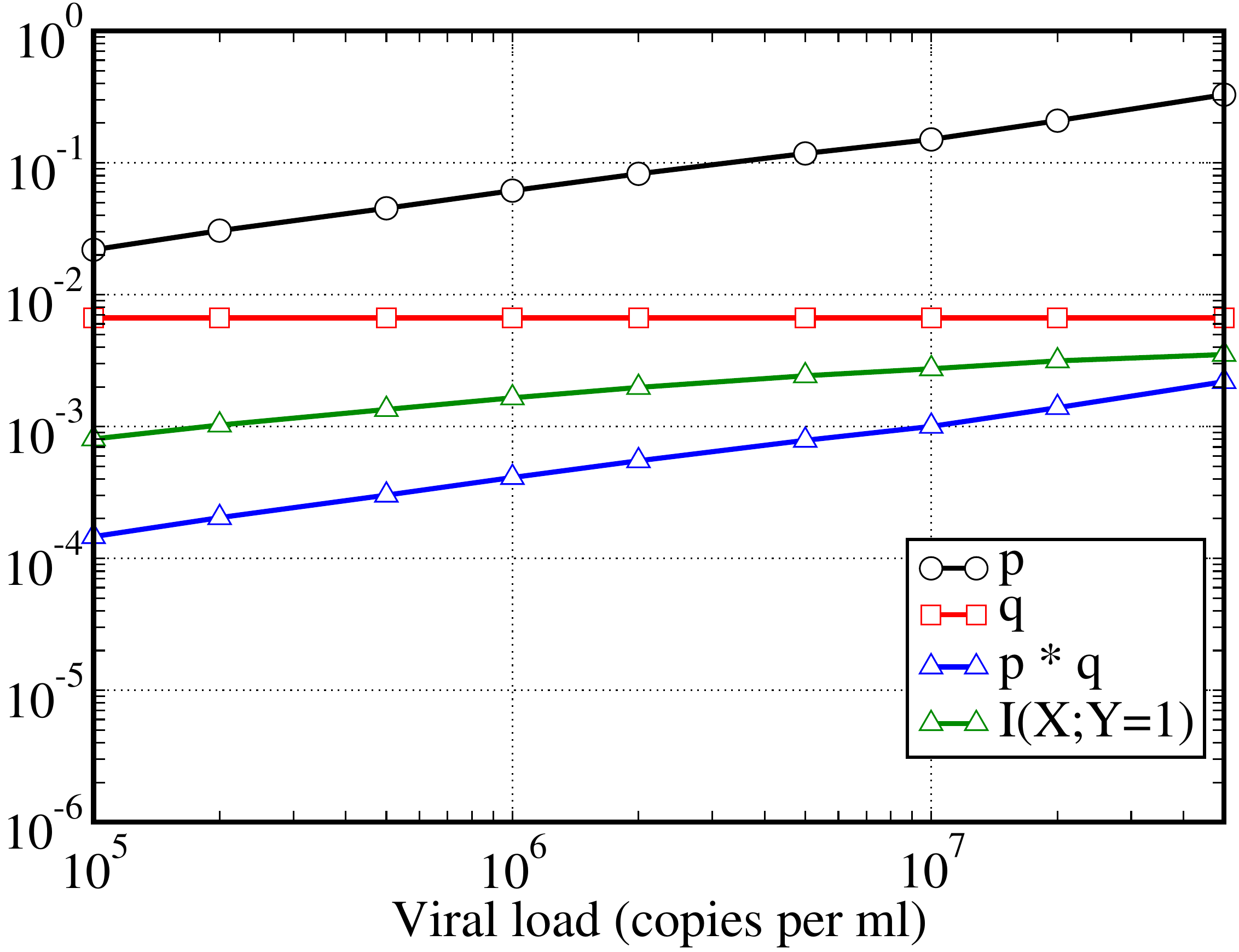}}
    \subfigure[Bus]{\includegraphics[width=0.49\linewidth]{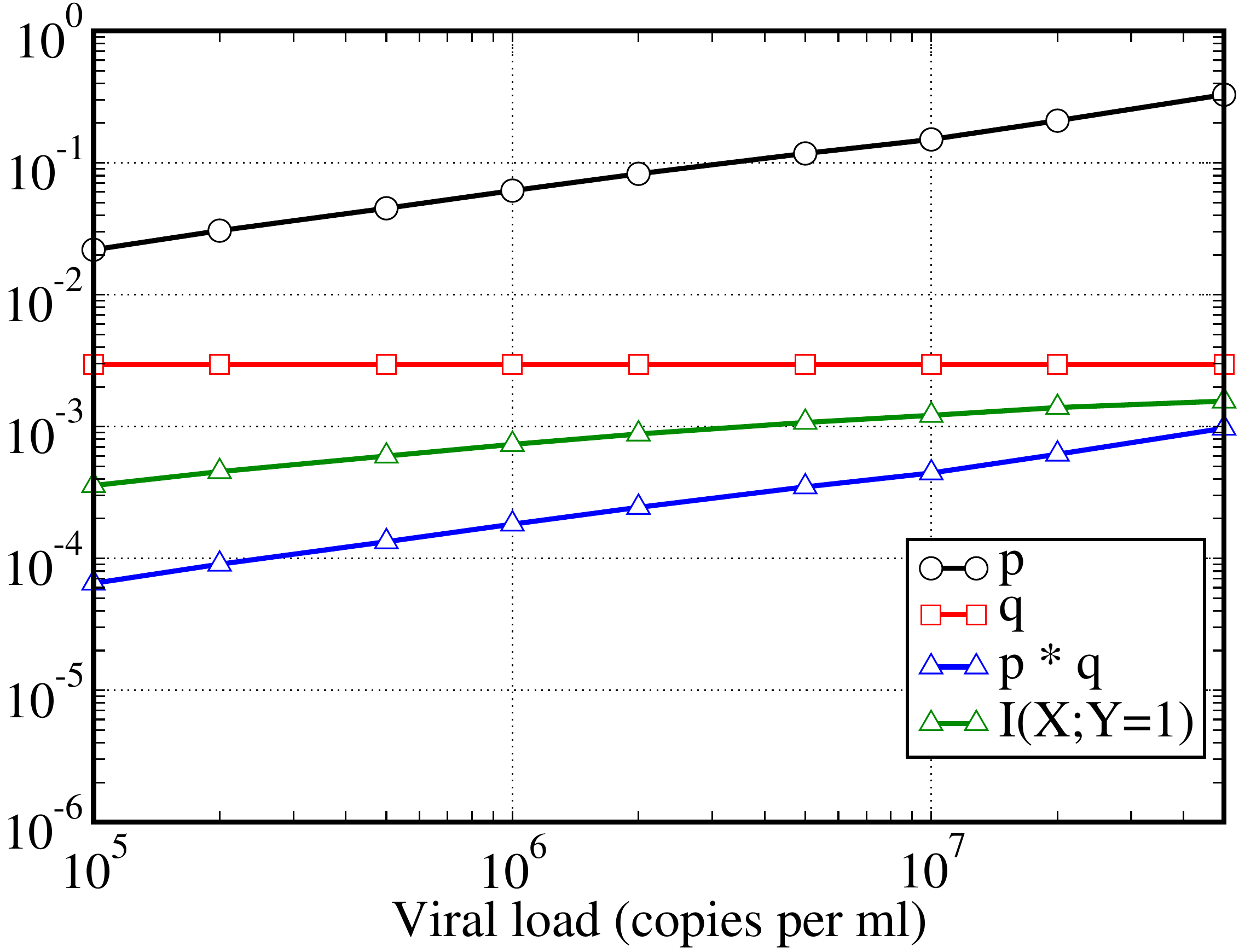}}
    \caption{Monte Carlo simulation results. Regarding classroom and bus, results are plotted only for the most critical receiver. The infection rate needs to be multiplied by the number of respiratory events, $n$.}%
    \label{fig:covsim}
    \vspace{-.8em}
\end{figure}

Numerical results obtained by Monte Carlo simulations are depicted in Fig.~\ref{fig:covsim}.  For each scenario, averaging is performed over 90-240 statistically independent runs. Besides the infection rate, the number of events, $n$, is an important parameter for the probability of infection.  Taking this into account, the office and the classroom are the most critical environments. 
Beyond a certain viral load the mutual information does not increase any more, which is counterintuitive. 

\section{Conclusions}
The so-called infection rate is a novel quantity to mathematically measure the ``infection performance.''  Besides a logarithmic infection rate based on the mutual information, an even simpler yet more realistic linear infection measure is suggested.
Unlike transmission schemes targeting a maximization of the mutual information, in the area of pathogen-laden particle transmission the objective is to minimize the infection rate. 
Both infection rates can completely be determined by the probability distribution of the channel input events, and by the transition probabilities between input and output events.  
This model separation proves to be useful. 
The input probabilities are affected by the kind of respiratory event, masks, etc. 
The transition probabilities are affected by distances, particle flow, and many other parameters. 
In this contribution, the mutual information has been calculated in closed form for several basic discrete memoryless channel models. 
It is shown that any additional infected transmitter linearly increases the infection rate. 
Furthermore, active and passive relaying are considered. 
By means of particle-based Monte Carlo simulations, the infection rates have been emulated for four typical environments: (a) two-person office space; (b) corridor; (c) classroom; (d) bus. 
Scenarios (a) and (c) are most critical. 


\bibliographystyle{sty/IEEEtranTCOM}
\bibliography{main}

\end{document}